\documentclass[11pt,a4paper]{article}

\usepackage[T1]{fontenc}
\usepackage[utf8]{inputenc}
\usepackage[margin=2.5cm]{geometry}
\usepackage{amsmath,amssymb,amsfonts}
\usepackage{bm}
\usepackage{graphicx}
\usepackage{booktabs}
\usepackage{array}
\usepackage{longtable}
\usepackage[numbers,sort&compress]{natbib}
\usepackage{xcolor}
\usepackage[hidelinks]{hyperref}
\usepackage{comment}

\newcommand{\doi}[1]{\href{https://doi.org/#1}{\textcolor{blue}{[DOI]}}}

\newcommand{\dd}{\mathrm{d}}
\newcommand{\Dt}{\mathrm{D}_t}
\newcommand{\Emf}{\mathcal{E}}
\newcommand{\Mlor}{\mathcal{M}}
\newcommand{\avg}[1]{\left\langle #1 \right\rangle}
\newcommand{\Rk}{\mathbf{R}}
\newcommand{\rk}{\mathbf{r}}
\renewcommand{\div}{\nabla\cdot}
\newcommand{\curl}{\nabla\times}
\newcommand{\jaumann}{\overset{\triangledown}{\mathrm{D}_t}}
\newcommand{\Iden}{\mathbf{I}}
\newcommand{\Stot}{\mathbf{t}}
\newcommand{\stf}[1]{\overline{#1}}
\newcommand{\Adev}{\stf{\nabla\mathbf{v}}}
\newcommand{\sPM}{\sigma_{\text{PM}}}
\newcommand{\sPMdev}{\stf{\sigma}_{\text{PM}}}

\title{Revisiting the Coupling of Thermodynamics and Electromagnetics}
\author{
  Stefanie Braun\\[1mm] 	
  {\small Applied and Computational Mathematics,}\\
  {\small RWTH Aachen University, Germany}
  \and
  Henning Struchtrup\\[1mm]
  {\small Mechanical Engineering,}\\
  {\small University of Victoria, Canada}
  \and
  Manuel Torrilhon\\[1mm]
  {\small Applied and Computational Mathematics,}\\
  {\small RWTH Aachen University, Germany}
}
\date{\today}

\begin{document}
\maketitle

\begin{abstract}
We revisit the coupling of continuum thermodynamics and electromagnetic theory for
polarisable and magnetisable matter in motion. Two routes are followed and then compared.
The first route is the axiomatic bulk theory of Dreyer, Guhlke and M\"uller, in which universal
balance laws are closed by an entropy principle. We show that the source of the internal energy
balance must be built with the non-convective electric current, that the polarisation current and
the Lorentz magnetisation enter through one single identity, which Dreyer et al.\ do not write
down, and that this identity fixes both the admissible entropy variables and the signs of
the bound-current ansatz. The second route is the statistical-mechanical one of Mazur, in
which the macroscopic Maxwell equations are obtained by ensemble averaging over a system of
atoms with internal charge carriers. Mazur stops before the conservation laws, so we derive them,
and we estimate the size of the mass-correction terms that appear. The comparison shows that
after a redefinition of polarisation and magnetisation the two sets of equations agree
structurally. The only irreducible difference is a momentum contribution from microscopic field
fluctuations, which can not be reproduced in a purely macroscopic theory. We further show that the
electromotive intensity $\Emf$ and the Lorentz magnetisation $\Mlor$ are not modelling choices
but appear by themselves, and that the asymmetric look of the entropy function is a consequence
of the chosen energy variable and not a defect of the theory.

\end{abstract}


\section{Introduction}\label{sec:intro}

In the recent decades, research areas involving charged systems of particles, e.g.\ within the
context of electromobility applications or medical studies of biochemical phenomena within the
body, have gained increased attention. A necessary prerequisite to study such systems is a
mathematical theory capable of describing thermodynamic systems of charged
particles, i.e., systems that are not only characterised by the thermodynamic parameters such
as mass, density, pressure etc.\ as they are standardly described in thermodynamic theory, but
also take properly into account the electromagnetic properties of the single particles and of
the system as a whole. Interestingly, although refined theories exist both for the
thermodynamic part (e.g., rational thermodynamics) as well as for the electromagnetic part
(Maxwell's theory of electromagnetics), a model that describes both aspects consistently seems
to be much harder to derive than is obvious at first glance.

Attempts to find such a model have been made especially in the field of electrochemistry,
where it became obvious that the traditionally used Nernst--Planck model suffered from severe
shortcomings, the most severe of which being that the resulting equations lead to negative
entropy production. Dreyer et al., in several papers
\cite{a1:dreyer2018,a1:dreyer2013,a1:dreyer2020}, have done pioneering work in this field,
ruling out the thermodynamic inconsistencies, yet there remain several issues on the
electromagnetic side, which are also present in other existing literature such as
\cite{a1:kovetz2006,a1:steigmann}. The main difficulties here lie in the consistent coupling of
the electromagnetic (EM) part, described by the Maxwell equations (which are field equations!),
to the thermodynamic part as described by rational thermodynamics.

One key aspect is to find a consistent macroscopic form for the EM equations that is applicable
to a system of moving charged particles, since the version of Maxwell's equations that is
standardly used in the literature is tailored to systems at rest or moving with uniform
velocity, but does not include the motion of the individual particles. Such a model can only
describe the effects of an externally applied EM field but cannot capture the dynamics of the
field that is created by the particles themselves and which is crucial to include in a
self-consistent theory.

The next challenge lies in finding the correct coupling terms in the conservation laws. Whereas the mass conservation
law remains the same as in a system of uncharged particles, EM effects do play a significant
role in the momentum and energy conservation laws when charged particles are considered.
Although separate conservation laws for both the kinetic part and the EM part can be found
rather easily, it is not a priori clear how they need to be coupled together. Different
versions exist in the literature, which all seem to suffer from one or the other shortcoming.

Once the equations are set up correctly, it is necessary to identify a fitting thermodynamic
potential (e.g., the Gibbs free energy) that leads to meaningful equations and sets the
variables for which constitutive laws have to be found. Again, a vast variety of different
versions exists in the literature, without a solution that clearly outstands the others and
leads to a contradiction-free theory.

Ultimately, the conservation laws, the coupling terms and any closure should ideally follow
from a microscopic theory. Mazur \cite{a1:mazur1957} comes closest to this. He derives the
macroscopic Maxwell equations from electron theory by ensemble averaging and he gives exact
microscopic formulas for polarisation, quadrupole and magnetisation. He does not, however, carry
the same averaging through to mass, momentum and energy, so the precise form of the conservation
laws, and with it the precise form of the coupling terms, is left open. Filling in this step is
one of the purposes of the present paper.

This paper revisits two routes to resolve the coupling and clarifies important aspects.
First, Dreyer, Guhlke and M\"uller \cite{a1:dreyer2018}
construct a general framework for magnetisable, polarisable, elastic, viscous, heat-conducting
and reactive mixtures. It rests on two pillars: universally valid balance equations, and an
axiomatic entropy principle from which the constitutive relations follow.  However, the choice of
variables, formulation of the energy law and the dependencies of the entropy function come with
significant questions which will be resolved below.
Second, Mazur \cite{a1:mazur1957} works from the single-charge-carrier picture. Below we
complete his framework by the conservation laws and discuss how an inconsistency of his theory
can be resolved.

The paper is organised as follows. Section~\ref{sec:setting} collects the classical conservation
laws, the Gibbs equation and the microscopic and macroscopic Maxwell equations, and states the
open questions. Section~\ref{sec:dreyer} goes through the bulk theory of Dreyer, Guhlke and
M\"uller: the energy balance, the polarisation current, the electro-magnetic Gibbs equation, the
entropy production and the sign ambiguity in the bound-current ansatz.
Section~\ref{sec:mazur} presents Mazur's statistical-mechanical route, derives the conservation
laws that are missing there and estimates the mass corrections. Section~\ref{sec:comparison}
compares the two sets of equations term by term and Section~\ref{sec:discussion} summarises what
follows and what stays open. The microscopic derivations are collected in
Appendix~\ref{app:A}.

\section{Thermodynamics and Electromagnetics}\label{sec:setting}

\subsection{Conservation laws and entropy for a single fluid}
\label{sec:classical-entropy}

The mass density $\rho$, the momentum density
$\rho\mathbf{v}$ and the total energy density $e_{\mathrm{tot}} = \rho u + \tfrac12\rho|\mathbf{v}|^2$
satisfy the three fundamental conservation laws
\begin{align}
  \partial_t\rho + \div(\rho\mathbf{v}) &= 0, \label{eq:mass1}\\
  \partial_t(\rho\mathbf{v}) + \div(\rho\mathbf{v}\mathbf{v} + \Stot) &= \mathbf{b},
  \label{eq:mom1}\\
  \partial_t\Big(\rho u + \tfrac12\rho v^2\Big)
  + \div\Big[\big(\rho u + \tfrac12\rho v^2\big)\mathbf{v} + \Stot\cdot\mathbf{v} + \mathbf{q}\Big]
  &= \mathbf{b}\cdot\mathbf{v} + r, \label{eq:energy1}
\end{align}

with the specific internal energy $u$, $v^2=|\mathbf{v}|^2$, the total stress tensor
$\Stot=(p+\pi)\,\Iden+\sigma$, containing the equilibrium pressure $p$ and
non-equilibrium pressure $\pi$,
as well as the deviatoric stress $\sigma$, a body force $\mathbf{b}$, the
heat flux $\mathbf{q}$ and an external supply $r$.

Using the material derivative the conservation laws for mass and energy can be re-written
as
\begin{align}
  \rho\, \Dt \upsilon \;-\; \div\mathbf{v} &= 0, \label{eq:mass}\\
  \rho\, \Dt u \;+\; \div\mathbf{q}
    &= -\,p\,\div\mathbf{v}
       \;-\; \sigma:\nabla\mathbf{v}-\; \pi\,\div\mathbf{v}, \label{eq:energy}
\end{align}
where $\upsilon=1/\rho$ is the specific volume. The second law states that these conservation laws imply an
entropy $s$ which satisfies the equations
\begin{align}
  \rho\, \Dt s \;+\; \div\bm{\varphi} &= \Sigma \label{eq:entropy}
\end{align}
where $\bm{\varphi}$ is the entropy flux and $\Sigma \ge 0$ is the entropy production.
We follow throughout the classical route of irreversible thermodynamics
\cite{a1:degrootmazur}.

In equilibrium or quasi-static processes the Gibbs equation effectively observes that $1/T$ acts as
integrating factor for the energy balance. We write the energy balance \eqref{eq:energy}
\begin{align}
 \rho\bigl(\Dt u + p\, \Dt \upsilon\bigr) &= -\div\mathbf{q}\;-\; \sigma:\nabla\mathbf{v}-\; \pi\,\div\mathbf{v}
\label{eq:energy2}
\end{align}
and the Gibbs equations becomes
\begin{equation}
  \frac{1}{T}\bigl(\Dt u + p\, \Dt \upsilon\bigr) \;=\; \Dt s,
  \qquad\text{i.e.}\qquad
  \frac{\partial s(u,\upsilon)}{\partial u} \;=\; \frac{1}{T},
  \qquad
  \frac{\partial s(u,\upsilon)}{\partial \upsilon} \;=\; \frac{p}{T}
  \label{eq:gibbs-relation}
\end{equation}
for the specific entropy $s = s(u,\upsilon)$. Simple transformations in equilibrium allows to derive
the usual dual forms
\begin{align}
  T\, \Dt s &= \Dt h \;-\; \upsilon\, \Dt p,
    &&h(s,p) = u + p\upsilon \label{eq:h-form}\\[2pt]
  \Dt f &= -\,s\, \Dt T \;-\; p\, \Dt \upsilon,
    &&f(T,\upsilon) = u - T s, \label{eq:f-form}\\[2pt]
  \Dt g &= -\,s\, \Dt T \;+\; \upsilon\, \Dt p,
    &&g(T,p) = h - T s, \label{eq:g-form}
\end{align}
which introduces enthalpy $h$, specific free energy $f$ and Gibbs free energy $g$. These provide many
different choices to describe equilibrium processes. For later reference, we remark that the
relations
\begin{equation}
  p(T,\upsilon) = -\frac{\partial f(T,\upsilon)}{\partial \upsilon},
  \qquad
  \upsilon(T,p) = \frac{\partial g(T,p)}{\partial p}.
  \label{eq:dual-equilibrium}
\end{equation}
both state equivalent constitutive relations between pressure and specific volume related through
Legendre transform.

In not-too-strong non-equilibrium the Gibbs equations can still be assumed to be valid and irreversible
thermodynamics deduces from \eqref{eq:energy2} after multiplication with $1/T$ the entropy flux
$\bm{\varphi}=\mathbf{q}/T$
and the production
\begin{equation}
  \Sigma = -\frac{1}{T^2}\mathbf{q}\cdot\nabla T - \frac{\pi}{T}\div \mathbf{v} - \frac{1}{T}\sigma : \nabla \mathbf{v} \ge 0
  \label{eq:entropy-production-nsf}
\end{equation}
which gives rise to the well-known closures of Fourier and Navier-Stokes.
The same structure will reappear below, only with more fluxes and more forces.

The continuum description contains an intrinsic coarse-graining. For a dilute gas the fields
\eqref{eq:mass}--\eqref{eq:energy} are the lowest moments of a one-particle distribution function
\cite{a1:chapmancowling}, and the closure problem for the higher moments is the kinetic analogue
of the thermodynamic closure above. The averaging that Mazur performs in
Section~\ref{sec:mazur} is of the same nature, only carried out over the phase space of atoms
with internal structure instead of over the velocity space of structureless particles.

\subsection{Microscopic and macroscopic Maxwell equations}

At the most fundamental level the electromagnetic field is governed by the microscopic Maxwell
equations, in SI units \cite{a1:jackson1999},
\begin{equation}
  \div\mathbf{b} = 0, \qquad
  \partial_t\mathbf{b} + \curl\mathbf{e} = 0, \qquad
  \div\mathbf{e} = \frac{n^{\mathrm{mic}}}{\varepsilon_0}, \qquad
  \varepsilon_0\partial_t\mathbf{e} - \frac{1}{\mu_0}\curl\mathbf{b} = -\mathbf{j}^{\mathrm{mic}},
  \label{eq:micro-maxwell}
\end{equation}
where $\mathbf{e}$ and $\mathbf{b}$ are the microscopic electric field and magnetic induction,
$n^{\mathrm{mic}}$ and $\mathbf{j}^{\mathrm{mic}}$ are the microscopic charge and current densities counting
every individual charge carrier, and $\varepsilon_0$, $\mu_0$ are the vacuum permittivity and
permeability with $c^2 = 1/(\varepsilon_0\mu_0)$.

In a thermodynamic system made of atoms and molecules, tracking every elementary charge is
neither feasible nor desirable. The idea is to average the fields $\mathbf{e}$ and $\mathbf{b}$
into macroscopic fields, $\mathbf{E}=\avg{\mathbf{e}}$ and $\mathbf{B}=\avg{\mathbf{b}}$, and to split
the averaged sources into a free and a bound part. The standard coarse-graining introduces the polarisation
$\mathbf{P}$ and the magnetisation $\mathbf{M}$ to account for the bound charges and currents
inside the atomic charge clouds. Lorentz \cite{a1:lorentz1909} used a spatial average over
``physically infinitesimal'' volumes. Mazur \cite{a1:mazur1957} and Mazur and Nijboer
\cite{a1:mazurnijboer1953} showed that the ensemble average is the statistically rigorous
alternative. For a medium at rest the auxiliary fields are then \emph{defined} as
\begin{equation}
  \mathbf{D} := \varepsilon_0\mathbf{E} + \mathbf{P}, \qquad
  \mathbf{H} := \tfrac{1}{\mu_0}\mathbf{B} - \mathbf{M},
  \label{eq:DH-rest}
\end{equation}
and the macroscopic Maxwell equations read
\begin{equation}
  \div\mathbf{B} = 0, \quad
  \partial_t\mathbf{B} + \curl\mathbf{E} = 0, \quad
  \div\mathbf{D} = n_F, \quad
  \partial_t\mathbf{D} - \curl\mathbf{H} = -\mathbf{j}_F,
  \label{eq:macro-maxwell}
\end{equation}
with the free charge density $n_F$ and the free current $\mathbf{j}_F$. However, writing the
Maxwell equations exclusively on the fundamental fields $\mathbf{E}$ and $\mathbf{B}$ is
possible,
\begin{equation}
  \div\mathbf{B} = 0, \quad
  \partial_t\mathbf{B} + \curl\mathbf{E} = 0, \quad
  \varepsilon_0\div\mathbf{E} = n_e, \quad
  \varepsilon_0\partial_t\mathbf{E} - \tfrac{1}{\mu_0}\curl\mathbf{B} = -\mathbf{j}_e,
  \label{eq:pure-maxwell}
\end{equation}
but must consider the total electric charges $n_e$ and total electric currents $\mathbf{j}_e$.
The two forms are equivalent, with $n_e=n_F+n_P$ and $\mathbf{j}_e=\mathbf{j}_F+\mathbf{j}_P$.
Which of them one prefers is only a question of how the material information is modelled.

The traditional approach treats the electromagnetic sub-system as external and couples the two
theories through the right-hand sides of the momentum and energy equations. The coupling terms
are the Lorentz force density and the Ohmic heating,
\begin{equation}
  \mathbf{b} \;=\; n_e\mathbf{E} + \mathbf{j}_e\times\mathbf{B},
  \qquad
  \mathbf{b}\cdot\mathbf{v} + r \;=\; \mathbf{j}_e\cdot\mathbf{E},
  \label{eq:naive-coupling}
\end{equation}

so that \eqref{eq:mom1} and \eqref{eq:energy1} obtain the additional sources, while the
stress and the heat flux may now contain electromagnetic contributions. It must be clarified
which charges, currents and fields these coupling terms are based on.

When $\mathbf{P}$ and $\mathbf{M}$ are neglected, no ambiguity arises and the total thermodynamic and
electromagnetic system can be easily closed formally by an entropy
principle, however, when polarization and magnetization are to be considered in the material difficulties arise:
The electromagnetic part has not been properly adapted to a moving thermodynamic medium.
The macroscopic Maxwell equations in the form \eqref{eq:macro-maxwell} with \eqref{eq:DH-rest}
were derived for media at rest or in uniform motion. Additionally, it remains unclear if and how those
additional material fields add to the coupling terms and crucially to energy and entropy.

\section{The Dreyer--Guhlke--M\"uller Framework}\label{sec:dreyer}

\subsection{Overview and modelling philosophy}

Dreyer et al.~set up an extensive paper \cite{a1:dreyer2018}, introducing the electromagnetic
quantities and also the derived equations from the beginning purely in the thermodynamic
picture. Accordingly, they obtain the balance equations for the free and bound currents purely
from exploiting charge conservation, without having to refer to the fourth Maxwell equation.
Whereas this ansatz leads to a  clear and self-consistent formulation, the drawback
is that it raises the question how polarisation and magnetisation, which are
created by inner-particle charge separation, enter this picture from a physical point of
view.

The theory rests on two pillars: universally valid balance equations for matter together with
Maxwell's equations, and an axiomatic entropy principle from which the constitutive relations
are derived. Galilean symmetry is postulated and constrains the transformation properties of
all fields. The original work covers bulk and surface, while we focus on resolving the questions
for the bulk equations.

\subsection{Energy balance}

One important finding of \cite{a1:dreyer2018} is that while the thermodynamic and electromagnetic
energy balances combine to a total energy balance when total charges and currents are considered,
the balance of internal energy reads
\begin{equation}
  \rho \Dt u + \div\mathbf{q}
  = -\big((p+\pi)\Iden+\sigma\big):\nabla\mathbf{v} + \mathbf{J}_e\cdot\Emf .
  \label{eq:internal-em}
\end{equation}
All Lorentz-force and ohmic heat contributions collect
into the single electromotive work term $\mathbf{J}_e\cdot\Emf$ with the Galilean-invariant electric field
\begin{equation}
\Emf = \mathbf{E}+\mathbf{v}\times\mathbf{B}.
\label{eq:Emf}
\end{equation}
Similarly, the electric current has a convective formulation
\begin{equation}
\mathbf{j}_e = n_e\mathbf{v} + \mathbf{J}_e,
\label{eq:je-split}
\end{equation}
which defines the non-convective current $\mathbf{J}_e$.
The naive product $\mathbf{J}_e\cdot\mathbf{E}$ is not the correct source, and neither is the
total product $\mathbf{j}_e\cdot\mathbf{E}$. Only the pairing of the non-convective current with the
electromotive intensity is invariant under a change of the observer, and only this pairing is left
over once the mechanical power of the Lorentz force has been subtracted from the total
electromagnetic power. Both factors matter here.
Only at rest one has $\mathbf{j}_e=\mathbf{J}_e$ and $\Emf=\mathbf{E}$.

We note that \eqref{eq:internal-em} can easily be embedded into the entropy theory of Sec.~\ref{sec:classical-entropy}
with no changes to Gibbs equation or the definition of entropy. The heat source simply
occurs in the entropy production as $\mathbf{J}_e\cdot\Emf / T$. The electric field, governed by the
Maxwell equations, is a field variable of the system while the current is considered an unknown
which requires a \emph{closure}. Equilibrium suggests $\mathbf{J}_e=0$, and in non-equilibrium
Ohm's law $\mathbf{J}_e = \eta\,\Emf$ with $\eta\ge0$ results in positive entropy production.

\subsection{Polarisation current and Lorentz magnetisation}

In a polarized and magnetized material the total current must be decomposed into a free (external) and a polarisation part,
$\mathbf{j}_e = \mathbf{j}_F + \mathbf{j}_P$, and in \cite{a1:dreyer2018} the polarisation charge and current are
introduced as the formal solution of the balance
$\partial_t n_P + \div(n_P\mathbf{v}+\mathbf{J}_P)=0$, namely
\begin{equation}
  n_P = -\div\mathbf{P}, \qquad
  \mathbf{j}_P = \partial_t\mathbf{P} + \curl\mathbf{M},
  \label{eq:kinematic-identity}
\end{equation}
We would like to point out another interpretation. The polarization current $\mathbf{j}_P$ remains the unknown for which
a closure is required in the energy balance \eqref{eq:internal-em}. Instead of something like an Ohm's law the above relation introduces two physical effects that
generate polarization current: the temporal change of a field $\mathbf{P}$ and the spatial change of another field $\mathbf{M}$.
Both fields are unknown and must be considered internal variables of the material which give rise to the current. These
fields replace the current as unknown and now require \emph{closure relations}. These relations could be local equilibrium
conditions or come as evolution equations. This point of view helps to choose variables and entropy dependencies below.

The authors of \cite{a1:dreyer2018} write \eqref{eq:kinematic-identity} with the Galilean-invariant Lorentz magnetisation
\begin{equation}
  \Mlor = \mathbf{M} + \mathbf{v}\times\mathbf{P}
  \label{eq:Mscript}
\end{equation}
in convective form. With standard vector identities and $\mathbf{j}_P = n_P\mathbf{v}+\mathbf{J}_P$ one obtains
\begin{equation}
  \Dt\mathbf{P} + \mathbf{P}(\div\mathbf{v}) - (\mathbf{P}\cdot\nabla)\mathbf{v}
  + \curl\Mlor = \mathbf{J}_P,
  \label{eq:jP-convective}
\end{equation}
which is written like an evolution for $\mathbf{P}$, but remains a definition of the current  $\mathbf{J}_P$. The evolution operator, however, is
identical to the one of the magnetic field when written with the Galilean-invariant electric field
\begin{equation}
  \Dt\mathbf{B} + \mathbf{B}(\div\mathbf{v}) - (\mathbf{B}\cdot\nabla)\mathbf{v}
  + \curl\Emf = 0.
  \label{eq:B-convective}
\end{equation}
Using the mass balance \eqref{eq:mass} in the form $\rho\Dt\hat{\mathbf{a}} = \Dt\mathbf{a}+\mathbf{a}\div\mathbf{v}$
for $\hat{\mathbf{a}}=\upsilon\mathbf{a}$, dotting \eqref{eq:jP-convective} with $\Emf$ and
subtracting \eqref{eq:B-convective} dotted with $\Mlor$ results in the identity
\begin{equation}
  \rho\Dt(\Mlor\cdot\hat{\mathbf{B}}) + \div(\Emf\times\Mlor) = -\sPM:\nabla\mathbf{v} + \rho\Emf\cdot \Dt \hat{\mathbf{P}} + \rho\hat{\mathbf{B}}\cdot \Dt\Mlor - \Emf\cdot\mathbf{J}_P
  \label{eq:MB-identity}
\end{equation}
which serves as the center-piece of the theory of \cite{a1:dreyer2018}, even though they never write it explicitly. To stay in specific quantities we
have introduced  $\hat{\mathbf{P}}=\upsilon\mathbf{P}$ and $\hat{\mathbf{B}}=\upsilon\mathbf{B}$ facilitating the mass balance and the quasi-stress tensor
$\sPM$ is given by
\begin{equation}
\sPM = \Emf\otimes\mathbf{P} - \Mlor\otimes\mathbf{B},
\label{eq:PM-tensor}
\end{equation}
which is non-symmetric. \eqref{eq:MB-identity} is an important
equation because it pinpoints the source of ambiguity and choice in the theory. First of all, though it looks like an evolution
equation for some energy, we would like to emphasise that the equation comes as an identity derived from merely the definition
of $\mathbf{J}_P$ and the Maxwell equation for $\mathbf{B}$ and carries no additional physical information. It can not even
be viewed as clear evolution equation because the time derivatives on the right-hand-side allow to change the dot-product term
on the left arbitrarily. For instance, we could also choose to add $\Dt(\Emf\cdot\hat{\mathbf{P}})$ after product rule.

Equation \eqref{eq:MB-identity} can be combined with the energy balance \eqref{eq:internal-em} to eliminate the heating term, hence, replace
the current $\mathbf{J}_P$ by the new unknown fields of polarization and magnetization.
Here and in the following we neglect free electric currents, so that
$\mathbf{J}_e=\mathbf{J}_P$ and no term $\mathbf{J}_F\cdot\Emf$ appears.
This gives the modified energy balance
\begin{equation}
  \rho \Dt (u + \Mlor\cdot\hat{\mathbf{B}}) + \div(\mathbf{q} + \Emf\times\Mlor)
  = -\left((p+\pi)\Iden+\sigma+\sPM\right):\nabla\mathbf{v} + \rho\Emf\cdot \Dt \hat{\mathbf{P}} + \rho\hat{\mathbf{B}}\cdot \Dt\Mlor 
  \label{eq:internal-modified}
\end{equation}
also provided in \cite{a1:dreyer2018}.

\subsection{The electro-magnetic Gibbs equation}

In the work of Dreyer et al.~no explicit discussion of the various forms of the Gibbs equation is given. It is well-known that in a quasi-static processes involving polarization and magnetization terms like $\mathbf{E}\cdot \dd\mathbf{P}$, $\mathbf{P}\cdot \dd\mathbf{E}$ or $\mathbf{B}\cdot \dd\mathbf{M}$ and $\mathbf{M}\cdot \dd\mathbf{B}$ change the energy of the system. Equation \eqref{eq:internal-modified} allows us to identify which energy belongs to what process. Similar to Sec.~\ref{sec:classical-entropy} we identify after re-arranging \eqref{eq:internal-modified}
\begin{equation}
  \frac{1}{T}\bigl(\Dt u + p_{\text{total}}\, \Dt \upsilon - \Emf\cdot \Dt \hat{\mathbf{P}} + \Mlor\cdot \Dt \hat{\mathbf{B}}\bigr) \;=\; \Dt s
   \label{eq:EM-gibbs-relation}
\end{equation}
with an equilibrium entropy $s(u,\upsilon, \hat{\mathbf{P}}, \hat{\mathbf{B}})$ where $u$ is the specific thermodynamic internal energy of the material without any contribution from electromagnetics and $p_{\text{total}}$ is the total pressure $p_{\text{total}}=p+\tfrac{1}{3}\operatorname{tr}(\sPM)$ with $p$ the thermodynamic pressure.
Which pressure and which energy belong into the electro-magnetic Gibbs relation has also been
discussed in Steigmann \cite{a1:steigmann} and the review of M\"uller et al.
\cite{a1:mueller2022}.

Using the usual techniques of Legendre-transforms and product rules it is possible to derive a collection of equivalent equilibrium statements. As in Sec.~\ref{sec:classical-entropy} we, for example, have the equivalent equilibrium closure relations
\begin{equation}
  \Emf(T,p,\hat{\mathbf{P}},\hat{\mathbf{B}}) = \frac{\partial g_1(T,p,\hat{\mathbf{P}},\hat{\mathbf{B}})}{\partial \hat{\mathbf{P}}},
  \qquad
  \hat{\mathbf{P}}(T,p,\Emf,\hat{\mathbf{B}}) = -\frac{\partial g_2(T,p,\Emf,\hat{\mathbf{B}})}{\partial \Emf}
  \label{eq:dual-equilibrium-EM}
\end{equation}
based on the Gibbs free energies
\begin{equation}
 g_1(T,p,\hat{\mathbf{P}},\hat{\mathbf{B}}) = u-T\,s+p\,\upsilon,
  \qquad
  g_2(T,p,\Emf,\hat{\mathbf{B}}) =  u-T\,s+p\,\upsilon-\Emf\cdot\hat{\mathbf{P}}.
  \label{eq:Gibbs-energies}
\end{equation}
Analogous relations can be found involving the magnetization. Note that the role of energy in the Gibbs equation for polarized and magnetized material is being debated. The derivation of \eqref{eq:EM-gibbs-relation} above goes through the fully inhomogeneous and possibly moving material to ensure consistency with the general situation and leaves no ambiguity.

\subsection{Entropy production and the Debye-type relaxation closure}

To derive the entropy law we split the velocity gradient
$\nabla \mathbf{v} = \Adev+\mathbf{W}+\tfrac{1}{3}\div\mathbf{v}\,\Iden$, where the
overbar denotes the symmetric, trace-free (deviatoric) part
and $\mathbf{W}$ is the antisymmetric part. Using the Jaumann derivative for vectors
\begin{equation}
\jaumann \mathbf{a} := \Dt \mathbf{a} - \mathbf{W}\cdot\mathbf{a}
  \label{eq:jaumann}
\end{equation}
we write the modified energy balance \eqref{eq:internal-modified} as \vspace{-1mm}
\begin{equation}
  \rho \Dt (u + \Mlor\cdot\hat{\mathbf{B}}) + \div(\mathbf{q} + \Emf\times\Mlor)
  =  -(p_{\text{total}}+\pi)\div\mathbf{v} - (\sigma+\sPMdev):\Adev +\rho\Emf\cdot \jaumann \hat{\mathbf{P}} + \rho\hat{\mathbf{B}}\cdot \jaumann\Mlor 
  \label{eq:internal-jaumann}
\end{equation}
where the overbar on $\sPMdev$ again marks the symmetric, trace-free part, here of the quasi-stress tensor \eqref{eq:PM-tensor}.
The antisymmetric part of $\sPM$ is not lost, it is precisely what turns the two
material derivatives on the right-hand side into Jaumann derivatives.
Note that the Jaumann derivative still satisfies a product rule 
$  \Dt(\mathbf{a}\cdot\mathbf{b})
  = \mathbf{a}\cdot\jaumann\mathbf{b} + \mathbf{b}\cdot\jaumann\mathbf{a}$.

The Gibbs equation \eqref{eq:EM-gibbs-relation} suggests to use an entropy with the
dependencies $s(u,\upsilon, \hat{\mathbf{P}}, \hat{\mathbf{B}})$, however, \cite{a1:dreyer2018} decides differently and lacks a detailed explanation.
We would phrase this decision as follows. The choice $s(u + \Mlor\cdot\hat{\mathbf{B}},\upsilon, \hat{\mathbf{P}}, \Mlor)$ as implied by \eqref{eq:internal-jaumann}, because of the time-derivatives on the right-hand-side, is clearly better suited to derive closure relations for our polarization and magnetization scenario.
The resulting entropy looks asymmetric in the electric and the magnetic variables, which has
been held against the theory. The asymmetry is however not a defect. It only records which energy
variable was chosen, and a different choice of energy variable moves the asymmetry to a different
place without changing any physics.
Indeed, we are looking for evolution equations for our unknown in the form\vspace{-1mm}
\begin{align}
  \rho\, \jaumann \hat{\mathbf{P}} &= \mathbf{r}^{P}(u,\upsilon, \hat{\mathbf{P}}, \Mlor, \Emf, \hat{\mathbf{B}}),
    \label{eq:P-evolution}\\
  \rho\, \jaumann \Mlor &= \mathbf{r}^{M}(u,\upsilon, \hat{\mathbf{P}}, \Mlor, \Emf, \hat{\mathbf{B}}),
    \label{eq:M-evolution}
 \end{align}
where the closure problem effectively asks for the relaxation source terms $\mathbf{r}^{P}$ and $\mathbf{r}^{M}$. These expressions can be used to substitute the time derivatives on the right hand side of \eqref{eq:internal-jaumann} and the source terms enter as thermodynamic fluxes into the entropy production. In particular, the evolutions of the electro-magnetic fields $\hat{\mathbf{B}}$ and $\Emf$ are fixed and using them in \eqref{eq:internal-jaumann} does not yield a useful entropy production.

When computing the evolution $\rho\,\Dt s(u_{\text{MB}},\upsilon, \hat{\mathbf{P}}, \Mlor)$ with $u_{\text{MB}}:=u + \Mlor\cdot\hat{\mathbf{B}}$ we use the equilibrium conditions implied by the corresponding Gibbs equation, in particular
\begin{align}
  -\frac{1}{T}\Emf_{\text{eq}}(u_{\text{MB}},\upsilon, \hat{\mathbf{P}}, \Mlor) = \frac{\partial s(u_{\text{MB}},\upsilon, \hat{\mathbf{P}}, \Mlor)}{\partial \hat{\mathbf{P}}},
 \label{eq:equilibrium-E} \\
  -\frac{1}{T}\hat{\mathbf{B}}_{\text{eq}}(u_{\text{MB}},\upsilon, \hat{\mathbf{P}}, \Mlor) = \frac{\partial s(u_{\text{MB}},\upsilon, \hat{\mathbf{P}}, \Mlor)}{\partial  \Mlor},
  \label{eq:equilibrium-B}
\end{align}
which yields the entropy  production
\begin{align}
\Sigma =& -\frac{1}{T^2}(\mathbf{q}+ \Emf\times\Mlor)\cdot\nabla T - \frac{1}{T}\pi\div \mathbf{v} - \frac{1}{T}(\sigma+\sPMdev) : \Adev\nonumber\\
 &\qquad\qquad\qquad+ \frac{1}{T}\rho\left(\Emf-\Emf_{\text{eq}}\right)\cdot \mathbf{r}^{P} + \frac{1}{T}\rho\left(\hat{\mathbf{B}}-\hat{\mathbf{B}}_{\text{eq}}\right)\cdot \mathbf{r}^{M}  \ge 0
  \label{eq:entropy-production-final}
\end{align}
which clarifies the expression in \cite{a1:dreyer2018} and allows a clean discussion of equilibrium and non-equilibrium closure.
The entropy flux is $\bm{\varphi}=(\mathbf{q}+\Emf\times\Mlor)/T$, so the electromagnetic
contribution to the entropy transport appears by itself as well.
Note that we have used the fact that $\partial s/\partial \hat{\mathbf{P}}\otimes\hat{\mathbf{P}}+\partial s/\partial \Mlor\otimes\Mlor$ is a symmetric tensor due to rotational invariance of $s$ to incorporate the antisymmetric velocity gradient into the Jaumann derivative of the equilibrium terms.

An equilibrium closure for polarization and magnetization sets $\Emf=\Emf_{\text{eq}}$ and $\hat{\mathbf{B}}=\hat{\mathbf{B}}_{\text{eq}}$. These are implicit conditions which require that $(\hat{\mathbf{P}},\Mlor)=(\hat{\mathbf{P}}_{\text{eq}},\Mlor_{\text{eq}})$ must be chosen such that these conditions hold for the electric and magnetic field according to \eqref{eq:equilibrium-E}/\eqref{eq:equilibrium-B}. No evolution equations for polarization and magnetization are necessary.

A non-equilibrium closure chooses a positive entropy production by setting, for instance
  \begin{equation}
  \mathbf{r}^{P}(\hat{\mathbf{P}}, \Mlor) = -\nu_P(\Emf_{\text{eq}}(\hat{\mathbf{P}}, \Mlor)-\Emf), \qquad \mathbf{r}^{M}(\hat{\mathbf{P}}, \Mlor)  = -\nu_M(\hat{\mathbf{B}}_{\text{eq}}(\hat{\mathbf{P}}, \Mlor)-\hat{\mathbf{B}})
  \label{eq:closure}
\end{equation}
with relaxation coefficients $\nu_P,\nu_M\ge0$. This is a Debye-type relaxation and it
can be used to close the evolution equations \eqref{eq:P-evolution}/\eqref{eq:M-evolution}. The concavity of the entropy $s$ ensures that the Jacobian of the right hand sides $( \mathbf{r}^{P}(\hat{\mathbf{P}}, \Mlor) , \mathbf{r}^{M}(\hat{\mathbf{P}}, \Mlor) )$ with respect to polarization and magnetization is negative definite, hence the relaxation is stable.
As mentioned above, changing the time derivatives in \eqref{eq:internal-jaumann} and then introducing a relaxation equation for, e.g., the magnetic field $\rho\, \jaumann \hat{\mathbf{B}} = \mathbf{r}^{B}$, as suggested in \cite{a1:dreyer2018}, is inconsistent with the fact that the magnetic field already follows from a Maxwell equation. No surprise, and quite satisfactory, Dreyer et al.~show that this leads to an instability. Note that, changing the time derivatives is certainly possible when the relevant evolution equations \eqref{eq:P-evolution}/\eqref{eq:M-evolution} are kept and possibly transformed into equations for different variables reflecting the dependencies of the new entropy. This will yield an equivalent expression for the production at the expense of much more involved terms.

\subsection{An ambiguity in the material theory}

As mentioned above the expression for the induced current $\mathbf{j}_P$ in \eqref{eq:kinematic-identity} can be viewed as an ansatz which introduces the internal variables
of polarization and magnetization. As such the signs of the various terms could be chosen differently. We introduce two factors $\lambda_{P,M}\in\{+1,-1\}$ and write
\begin{equation}
  n_P = -\lambda_P \div\mathbf{P}, \qquad
  \mathbf{j}_P = \lambda_P\partial_t\mathbf{P} + \lambda_M\curl\mathbf{M},
  \label{eq:kinematic-identity-sign}
\end{equation}
such that charge conservation is still satisfied. This ambiguity has been overlooked in \cite{a1:dreyer2018}. We can define $\Mlor = \mathbf{M} + \lambda_P\lambda_M\mathbf{v}\times\mathbf{P}$ and arrive at
\begin{equation}
  \lambda_P\left(\Dt\mathbf{P} + \mathbf{P}(\div\mathbf{v}) - (\mathbf{P}\cdot\nabla)\mathbf{v}\right)
  + \lambda_M\curl\Mlor = \mathbf{J}_P.
  \label{eq:jP-convective-sign}
\end{equation}
These signs can be carried through the entire calculation. Note, that we keep the postulation of the evolution equations \eqref{eq:P-evolution}/\eqref{eq:M-evolution} unchanged. The signs finally occur in the entropy production such that we have
\begin{equation}
 \frac{1}{T}\rho\left(\lambda_P\Emf-\Emf_{\text{eq}}\right)\cdot \mathbf{r}^{P} + \frac{1}{T}\rho\left(\lambda_M\hat{\mathbf{B}}-\hat{\mathbf{B}}_{\text{eq}}\right)\cdot \mathbf{r}^{M}
 \label{eq:production-sign}
\end{equation}
in \eqref{eq:entropy-production-final}.
The equilibrium condition demands $\lambda_P\Emf=\Emf_{\text{eq}}$ and
$\lambda_M\hat{\mathbf{B}}=\hat{\mathbf{B}}_{\text{eq}}$, with
$\Emf_{\text{eq}}=-T\,\partial s/\partial\hat{\mathbf{P}}$ and
$\hat{\mathbf{B}}_{\text{eq}}=-T\,\partial s/\partial\Mlor$ from
\eqref{eq:equilibrium-E}/\eqref{eq:equilibrium-B}. Concavity of $s$ makes both equilibrium fields
increasing functions of their own variable, so a negative $\lambda$ would force a material that
polarises against the applied field. Together with a positive susceptibility the second law
therefore fixes $\lambda_P=\lambda_M=1$.
The coupling structure between $(\mathbf{P},\Mlor)$ and $\mathbf{J}_P$ is therefore an
output of the second law and not an input of the model. An ansatz with the wrong sign in either
term is thermodynamically inadmissible.

\section{Mazur's Statistical-Mechanical Theory}\label{sec:mazur}

\subsection{Overview and motivation}

In his paper \cite{a1:mazur1957}, Mazur derives the Maxwell equations for a thermodynamic
system of atoms or molecules (molecules can be treated within a straightforward generalisation
of the version that Mazur describes), starting from a single-particle picture on the level of
the individual charge carriers. With the help of an ensemble-averaging procedure it is possible
to coarse-grain the equations obtained on the single-charge-carrier level to the
``macroscopic'' particle level of atoms or molecules moving as a unit in a thermodynamic
system, keeping thorough track of all contributions from the motion of the single charge
carriers. Thus it is possible to make exact definitions of the electric and magnetic moments
and to keep track of how they enter the macroscopic equations.

In particular, from this derivation it becomes obvious that if magnetisation effects shall be
included in the theory, it is consistent to retain also the electric quadrupole moment, since
it is of the same order of magnitude regarding an ordering in the smallness of the deviation of
the charge carriers from the centre of gravity of the atom or molecule. Mazur's focus is on the
EM equations, and whereas he shows several examples for specific simple systems, his treatment
does not include a general thermodynamic framework for the conservation equations. The
derivation of these equations from Mazur's ansatz is done in Appendix~\ref{app:A}; here we only
state the results.

\subsection{Microscopic setup}

We follow the formalism introduced by Mazur and want to describe a thermodynamic system of
polarisable/magnetisable and/or charged particles. Let us consider a system of $N$ atoms or
ions, indexed with $k$; a generalisation to molecules is straightforward. Each atom consists of
$n_k$ individual charge carriers (electrons and the nucleus) with mass $m_{ki}$ and charge
$e_{ki}$, $k=1,\dots,N$, $i=1,\dots,n_k$. The position of each charge carrier is denoted by
$\Rk_{ki}$.

This notation, differentiating into individual charge carriers, is needed in the EM equations,
where the exact position of each charge plays a role, so that the microscopic structure is
important. On the macroscopic length scale, on which thermodynamic processes such as convection
occur, only the motion of the particles as a whole is of interest. On this length scale we can
group together the single charge carriers to an atom or ion and describe the motion of the
whole atom/ion instead. The $k$-th atom/ion is described by its centre of gravity
\begin{equation}
  \Rk_k = \textstyle \sum_i m_{ki}\Rk_{ki}/m_k ,
  \label{eq:centre-of-gravity}
\end{equation}
and the total mass and total charge of a particle are
\begin{equation}
  m_k = \textstyle \sum_i m_{ki}, \qquad e_k = \textstyle \sum_i e_{ki},
  \label{eq:mk-ek}
\end{equation}
respectively. The positions of the charge carriers within the particles are then described
relative to their centre of gravity as
\begin{equation}
  \rk_{ki} = \Rk_{ki} - \Rk_k ,
  \qquad\text{so that}\qquad
   \textstyle\sum_i m_{ki}\rk_{ki} = 0 .
  \label{eq:rki}
\end{equation}
Differentiating \eqref{eq:rki} in time also gives $\sum_i m_{ki}\dot\rk_{ki}=0$, which will be
used repeatedly below.
We introduce a probability distribution function
\begin{equation}
  f = f\big(\Rk_k^N,\ \rk_{ki}^{N(n_k-1)},\ \mathbf{p}_k^N,\ \mathbf{p}_{ki}^{N(n_k-1)},\ t\big),
  \label{eq:pdf}
\end{equation}
and define the average over the whole phase space as
\begin{equation}
  \avg{\,\cdot\,} \equiv \int \dd\Rk_k^N\,\dd\rk_{ki}^{N(n_k-1)}\,
  \dd\mathbf{p}_k^N\,\dd\mathbf{p}_{ki}^{N(n_k-1)}\ (\cdot),
  \qquad \avg{f} = 1 .
  \label{eq:ensemble-average}
\end{equation}
Due to the conservation of probability in phase space we know that
\begin{equation}
  \frac{\dd f}{\dd t} = \frac{\partial f}{\partial t}
  + \sum_k\Big[(\dot\Rk_k\cdot\nabla_{\Rk_k})f + (\dot\rk_{ki}\cdot\nabla_{\rk_{ki}})f
  + (\dot{\mathbf{p}}_k\cdot\nabla_{\mathbf{p}_k})f
  + (\dot{\mathbf{p}}_{ki}\cdot\nabla_{\mathbf{p}_{ki}})f\Big] = 0 .
  \label{eq:liouville}
\end{equation}
Equation \eqref{eq:liouville} is the starting point for everything that follows.

\subsection{\texorpdfstring{Ensemble-averaged definitions of $\mathbf{P}$, $\mathbf{Q}$ and
  $\mathbf{M}$}{Ensemble-averaged definitions of P, Q and M}}

We introduce the electric and magnetic moments of the particle $k$,
\begin{equation}
  e_k = \textstyle \sum_i e_{ki},\qquad
  \bm{\mu}^{\mathrm{el}}_k = \textstyle \sum_i e_{ki}\rk_{ki},\qquad
  \mathbf{Q}_k = \textstyle \frac12\sum_i e_{ki}\rk_{ki}\rk_{ki},\qquad
  \bm{\mu}^{m}_k = \textstyle \frac12\sum_i e_{ki}\,\rk_{ki}\times\dot\rk_{ki},
  \label{eq:moments}
\end{equation}
which represent the charge, electric dipole and quadrupole moments and the magnetic dipole moment,
respectively. These moments are defined with respect to the corresponding centre of gravity as
origin. Summing over all particles and taking the ensemble average gives the polarisation
$\mathbf{P}$, the quadrupole $\mathbf{Q}$ and the magnetisation $\mathbf{M}$,
\begin{align}
  \mathbf{P}(\mathbf{R},t) &= \Big\langle \textstyle \sum_{k} \bm{\mu}^{\mathrm{el}}_k\,
      \delta(\Rk_k-\mathbf{R})f\Big\rangle, \label{eq:Pdef}\\
  \mathbf{Q}(\mathbf{R},t) &= \Big\langle \textstyle\sum_{k}  \mathbf{Q}_k\,
      \delta(\Rk_k-\mathbf{R})f\Big\rangle, \label{eq:Qdef}\\
  \mathbf{M}(\mathbf{R},t) &= \Big\langle \textstyle\sum_{k}  \bm{\mu}^{m}_k \,
      \delta(\Rk_k-\mathbf{R})f\Big\rangle. \label{eq:Mdef}
\end{align}
It is essential that these are exact definitions in terms of the electronic structure. Nothing
is postulated here; $\mathbf{P}$, $\mathbf{Q}$ and $\mathbf{M}$ are given quantities once the
microscopic model is given. It is, however, important to note that these macroscopic quantities are averaged over
the molecules at position $\Rk_k$ such that the sub-particle distribution only contributes through
the moments. This distinction becomes also relevant for the conservation laws below.

\subsection{The macroscopic Maxwell equations}

Averaging the microscopic Maxwell equations \eqref{eq:micro-maxwell} over the phase space gives rise
to the total current and the total charge
\begin{equation}
  \mathbf{j}_e \equiv \Big\langle \textstyle\sum_{k,i}e_{ki}\dot\Rk_{ki}\delta(\Rk_{ki}-\mathbf{R})f\Big\rangle,
  \qquad
  n_e \equiv \Big\langle \textstyle\sum_{k,i}e_{ki}\delta(\Rk_{ki}-\mathbf{R})f\Big\rangle .
  \label{eq:je-ne-mazur}
\end{equation}
These quantities split into a free and a bound part, $\mathbf{j}_e=\mathbf{j}_F+\mathbf{j}_P$,
$n_e = n_F + n_P$, where the free part simply averages the charges of the molecules while
the polarization part is obtained by expanding the delta function around $\Rk_k$ up to second
order in $|\rk_{ki}|/|\Rk_k|$. The result is
\begin{align}
  n_P &= -\div\big(\mathbf{P}-\div\mathbf{Q}\big) , \label{eq:nP-mazur}\\
  \mathbf{j}_P &= \partial_t(\mathbf{P}-\div\mathbf{Q})
   + \curl\big(\mathbf{M} - \mathbf{v}\times(\mathbf{P}-\div\mathbf{Q})\big) \nonumber\\
  &\qquad - \curl\Big\langle \textstyle\sum_k\mathbf{v}_k'\times
    \big(\bm{\mu}^{\mathrm{el}}_k-\div\mathbf{Q}_k\big)
    \delta(\Rk_k-\mathbf{R})f\Big\rangle ,
  \label{eq:jP-mazur}
\end{align}
where the velocity has been split into an averaged and a fluctuating part,
\begin{equation}
  \mathbf{v}_k' = \dot\Rk_k - \mathbf{v} .
  \label{eq:vkprime}
\end{equation}
After inserting these expressions into the Maxwell equations we can identify
\begin{align}
  \mathbf{D} &\equiv \varepsilon_0\mathbf{E} + \mathbf{P} - \div\mathbf{Q},
  \label{eq:Dmazur}\\
  \mathbf{H} &\equiv \tfrac{1}{\mu_0}\mathbf{B} - \mathbf{M}
   + \mathbf{v}\times(\mathbf{P}-\div\mathbf{Q})
   + \Big\langle\sum_k \mathbf{v}_k'\times\big(\bm{\mu}^{\mathrm{el}}_k-\div\mathbf{Q}_k\big)
     \delta(\Rk_k-\mathbf{R})f\Big\rangle
  \label{eq:Hmazur}
\end{align}
and we arrive at the macroscopic Maxwell equations \eqref{eq:macro-maxwell}.
Two improvements over the classical relations \eqref{eq:DH-rest} are visible. First,
$\mathbf{D}$ contains the quadrupole correction $-\div\mathbf{Q}$. Second, $\mathbf{H}$
accounts for the contributions of all moving polarised particles, not only for a medium at rest
or in uniform motion. Both corrections are of the same formal expansion order in
$|\rk_{ki}|/|\Rk_k|$ as
the magnetisation itself, so that dropping them while keeping $\mathbf{M}$ would be
inconsistent.

\subsection{Evolution of the polarisation and the role of the quadrupole}

Since the charge carriers all remain fixed to a specific atom or ion, the free and the total
charge are each conserved, and from the definitions one derives
\begin{equation}
  \partial_t n_F = -\div\mathbf{j}_F, \qquad
  \partial_t n_e = -\div\mathbf{j}_e ,
  \label{eq:charge-conservation-mazur}
\end{equation}
so that the bound charge is conserved as well. The corresponding balance is Mazur's equation
(I.28), which is \eqref{eq:nP-mazur}/\eqref{eq:jP-mazur} above.
The relation \eqref{eq:jP-mazur} is an exact
lab-frame balance for the bound charge, containing convection, the diffusive transport of
dipoles by the fluctuating velocities $\mathbf{v}_k'$, and the local change of the dipole
density. It is \emph{not} a constitutive law: the magnetisation enters only one level later,
through the relation between the rate $\dot{\mathbf{Q}}_k$ and $\mathbf{M}$. Mazur
stops here.
He does not impose any frame-indifferent closure, and this is exactly the point where the two
theories part ways. We return to it in Section~\ref{sec:comparison}.

\subsection{Derivation of the Conservation Laws}

The paper \cite{a1:mazur1957} by Mazur focuses on the ponderomotive force, that is, the force exerted by an
electromagnetic field on a polarisable dielectric. However, the framework allows to also derive the conservation laws
from the microscopic setup. Appendix \ref{app:A} gives the details while here only the results are discussed.

\subsubsection{Continuity equation}

Appendix \ref{app:A1} introduces particle and molecular masses and momenta.
The macroscopic density and velocity can be defined on the charge-carrier level,
as $\avg{\sum_{k,i}m_{ki}\delta(\Rk_{ki}-\mathbf{R})f}$, or on the molecular level,
$\rho \equiv \avg{\sum_{k}m_{k}\delta(\Rk_{k}-\mathbf{R})f}$ which follows the definition of 
the electro-magnetic quantities in \eqref{eq:Pdef}--\eqref{eq:Mdef}. Following Mazur we average on the charge-carrier
level to derive evolution equations, but identify variables on the molecular level and carry the difference explicitly.
The result reads
\begin{multline}
  \partial_t\rho
  + \partial_t\Big\langle\sum_{k,i}\tfrac12 m_{ki}\rk_{ki}\rk_{ki}:\nabla\nabla\,
    \delta(\Rk_k-\mathbf{R})f\Big\rangle \\
  + \div\Big[\rho\mathbf{v}
   + \Big\langle\sum_{k,i}\Big(\tfrac12 m_{ki}\dot\Rk_k\rk_{ki}\rk_{ki}:\nabla\nabla
   - m_{ki}\dot\rk_{ki}(\rk_{ki}\cdot\nabla)\Big)\delta(\Rk_k-\mathbf{R})f\Big\rangle\Big] = 0,
  \label{eq:continuity-full}
\end{multline}
which carries correction terms. Neglecting those second-order
mass corrections, see Sec.~\ref{sec:mass-estimate}, we arrive at the usual continuity equation
$\partial_t\rho + \div(\rho\mathbf{v}) = 0$.

\subsubsection{Momentum equation}

For the momentum balance we average $m_{ki}\dot\Rk_{ki}$ which brings in the electric and magnetic fields
into the equation through Newton's law of motion. The final equation contains averages of the
field fluctuations $\mathbf{e}' = \mathbf{e}-\avg{\mathbf{e}}$ and $\mathbf{b}' = \mathbf{b}-\avg{\mathbf{b}}$.
The final averaged momentum balance reads
\begin{multline}
  \partial_t\big(\rho\mathbf{v}+\varepsilon_0\mathbf{E}\times\mathbf{B}\big)
  + \partial_t\avg{\varepsilon_0\mathbf{e}'\times\mathbf{b}'f}
  + \partial_t\Big\langle\sum_{k,i}\Big(\tfrac12 m_{ki}\dot\Rk_k\rk_{ki}\rk_{ki}:\nabla\nabla
   - m_{ki}\dot\rk_{ki}(\rk_{ki}\cdot\nabla)\Big)\delta f\Big\rangle \\
  + \div\Big[\rho\mathbf{v}\mathbf{v}
  + \frac12\Big(\varepsilon_0\mathbf{E}^2+\frac{1}{\mu_0}\mathbf{B}^2\Big)\Iden
  - \Big(\varepsilon_0\mathbf{E}\mathbf{E}+\frac{1}{\mu_0}\mathbf{B}\mathbf{B}\Big)
  {}+ \Stot\Big] = 0 ,
  \label{eq:momentum-full}
\end{multline}
where the total stress tensor $\Stot$, with the sign convention of \eqref{eq:mom1}, now contains contributions from the mass corrections and the fluctuations,
see Appendix \ref{app:A3}.

\subsubsection{Energy equation}
Finally the energy balance results from averaging $\tfrac12 m_{ki}\dot\Rk_{ki}^2$ which again brings
in energetic contributions from the electromagnetic fields. Interestingly, the structure of the equation and
its variables allows to move any perturbation through mass corrections or field fluctuations into the
respective variables. This yields the final total energy conservation in the canonical form
\begin{equation}
  \partial_t\Big[\tfrac12\rho v^2 + u_{\mathrm{int}}
   + \frac12\Big(\varepsilon_0\mathbf{E}^2+\frac{1}{\mu_0}\mathbf{B}^2\Big)\Big]
  + \div\Big[\big(\tfrac12\rho v^2 + u_{\mathrm{int}}\big)\mathbf{v}
   + \mathbf{q} + \frac{1}{\mu_0}\mathbf{E}\times\mathbf{B}\Big] = 0,
  \label{eq:energy-full}
\end{equation}
where both the internal energy and heat flux contain non-trivial expressions, see
Appendix \ref{app:A4} and \eqref{eq:uint-micro} in particular.

\subsection{Estimate of the mass-correction terms}\label{sec:mass-estimate}

In the conservation equations there appear mass-correction terms of second order in
$|\rk_{ki}|/|\Rk_k|$. These stem from the fact that the mass of the particles is counted at the
centre of gravity instead of at the places of the individual charge carriers, which leads to a
slight offset. In practice, since the distance between the charge carriers and the centre of
gravity of the atom is very small compared with the macroscopic distances within the
thermodynamic system, these second-order terms can safely be neglected in a purely
thermodynamic system.

However, if we look at a system of charged particles and want to include EM effects due to
polarisation and magnetisation, it turns out that the magnetisation terms are formally also of
second order in $|\rk_{ki}|/|\Rk_k|$, so that, if we employ this ordering scheme and truncate
after the linear term, we cannot capture magnetisation effects. We therefore need a different
argument, in which the ordering is taken up to second order but the mass corrections can still
be dropped against the magnetisation terms.

Most of the mass-correction terms can be incorporated into $\Stot$, $\mathbf{q}$ and
$u_{\mathrm{int}}$. The only terms that remain and cannot structurally be put into another
quantity are those in the flux of the continuity equation and in the momentum density; these
two coincide. We therefore compare their magnitude with the magnetisation terms as they occur
in the momentum equation:
\begin{equation}
\begin{aligned}
  \frac{\displaystyle\sum_{k,i}\Big(\tfrac12 m_{ki}\dot\Rk_k\rk_{ki}\rk_{ki}:\nabla\nabla
   - m_{ki}\dot\rk_{ki}(\rk_{ki}\cdot\nabla)\Big)\delta(\Rk_k-\mathbf{R})}
  {\displaystyle\sum_{k,i}e_{ki}(\rk_{ki}\times\mathbf{b})(\rk_{ki}\cdot\nabla)
   \delta(\Rk_k-\mathbf{R})}
  \ &\sim\ \frac{m_{ki}\,|\Rk_k|\,r^2/(t\,L^2)}{e_{ki}|\mathbf{b}|\,r^2/L}
\end{aligned}
  \label{eq:mass-estimate}
\end{equation}
which can be simplified to
\begin{equation}
  \frac{m_{ki}\,|\Rk_k|}{e_{ki}|\mathbf{b}|\,t\,L}
  \ \sim\ \Big|\frac{1}{\Omega_{c,ki}\,t}\Big| .
  \label{eq:mass-estimate-2}
\end{equation}

This is done for the first term in the numerator; a similar estimate with the same result holds
for the second one. Here $r=|\rk_{ki}|$ and $1/L$ is the length scale of the nabla operator, which we order to be
on the thermodynamic length scale, i.e.\ in the same range of magnitude as $|\Rk_k|$, and $t$ is ordered on the time scale on which the thermodynamic processes occur.
The intra-particle length $r$ cancels between numerator and denominator, and with
$|\Rk_k|\sim L$ the last step follows.
The frequency
\begin{equation}
  \Omega_{c,ki} = \frac{e_{ki}|\mathbf{b}|}{m_{ki}}
  \label{eq:cyclotron}
\end{equation}
is the cyclotron frequency, i.e.\ typically the frequency with which charged particles gyrate
around a magnetic field line. For a magnetic field strength of $1\,$T it is in the MHz range for
ions and even in the GHz range for electrons. Thus, even for much smaller field strengths, it
seems safe to neglect the second-order mass corrections against the second-order EM terms which
we want to keep, since the thermodynamic processes we are interested in happen on much slower
time scales than the gyration of the electrons and ions.

\section{Comparison}\label{sec:comparison}

Table~\ref{tab:comparison} puts the two sets of equations next to each other. In the following
subsections we go through the entries and discuss where they agree, where they differ, and why.

\begin{table}[htbp]
\centering
\small
\begin{tabular}{>{\raggedright\arraybackslash}p{0.30\textwidth}
                >{\raggedright\arraybackslash}p{0.29\textwidth}
                >{\raggedright\arraybackslash}p{0.29\textwidth}}
\toprule
 & \textbf{Dreyer et al.} (macroscopic) & \textbf{Mazur-based} (microscopic) \\
\midrule
Continuity
 & $\partial_t\rho+\div(\rho\mathbf{v})=0$
 & same, plus second-order mass corrections \\[1mm]
Momentum
 & $\partial_t(\rho\mathbf{v}+\varepsilon_0\mathbf{E}\times\mathbf{B})
    +\div(\rho\mathbf{v}\mathbf{v}{}+\Stot)=\mathbf{b}$
 & same, plus $\partial_t\avg{\varepsilon_0\mathbf{e}'\times\mathbf{b}'f}$
   and mass corrections \\[1mm]
Energy
 & $\mathbf{J}_e\cdot\Emf$ as source
 & same after regrouping; differences go into $u_{\mathrm{int}}$ and $\mathbf{q}$ \\[1mm]
Charge / currents
 & formal solution of charge conservation
 & explicit ensemble averages; \emph{structurally identical} \\[1mm]
Displacement
 & $\mathbf{D}=\varepsilon_0\mathbf{E}+\mathbf{P}$
 & $\mathbf{D}=\varepsilon_0\mathbf{E}+\mathbf{P}-\div\mathbf{Q}$ \\[1mm]
Magnetic field
 & $\mathbf{H}=\mathbf{B}/\mu_0-\mathbf{M}$
 & $\mathbf{H}=\mathbf{B}/\mu_0-\mathbf{M}+\mathbf{v}\times(\mathbf{P}-\div\mathbf{Q})+\dots$ \\[1mm]
Polarisation balance
 & kinematic identity \eqref{eq:kinematic-identity}
 & exact lab-frame balance \eqref{eq:jP-mazur} \\[1mm]
Closure for $\mathbf{P}$
 & Debye-type relaxation \eqref{eq:closure}
 & none (left as microscopic fluxes) \\
\bottomrule
\end{tabular}
\caption{The two sets of equations of Dreyer et al.~and Mazur at a glance.}
\label{tab:comparison}
\end{table}

\subsection{Where the two theories agree}

We see that the continuity equations differ only by the second-order mass corrections in the
Mazur version, which can be neglected as shown above. Thus the two
continuity equations coincide. The same second-order mass corrections appear in the momentum
equation and can be neglected there as well.

There are two additional structurally independent differences between the two versions of the
momentum equation. In the Mazur convection term there appear terms containing
$\mathbf{v}_k'$, which stem from the deviations of the single-particle velocities from the mean
velocity $\mathbf{v}$. These contributions are not taken into account in the Dreyer version.
However, structurally they can be put into the stress tensor $\Stot$, so as long as $\Stot$ is
not determined in the Dreyer version, there is no formal contradiction. The second structural
difference stems from the deviations of the $\mathbf{e}$ and $\mathbf{b}$ fields from their
averaged versions $\mathbf{E}$ and $\mathbf{B}$. Since these appear both in the convection part
and in the time derivative, the equation cannot be formally put into the same form without
disregarding the field fluctuations. This means that there is a genuine structural difference
between the two momentum equations, stemming from field fluctuations on the microscopic
scale. If those fluctuations are neglected, the two equations coincide again.

For the energy equations, the expression for the heat source in Dreyer's version can be converted into
structurally the same field terms as in the Mazur-based version. The differences can again be
split into second-order mass corrections and deviations of the electromagnetic fields from their averages.
Structurally, all these differences can be put either into the internal energy or into the heat
flux, so that the two versions coincide.

For the electromagnetic equations that do not involve the polarisation or the magnetisation
 all equations coincide exactly, and here there are no differences due to the
deviations of the electromagnetic fields from their averages. Whereas Mazur's equations result from an
explicit calculation of the Maxwell equations based on the electronic structure of the
material, Dreyer obtains the expressions for the currents from a formal solution of the charge
conservation equations in fluid form and defines the total charge density via the electric
field, which in Mazur's setting is the first Maxwell equation.

\subsection{Matching polarisation and magnetisation}

The equations containing $\mathbf{P}$ and $\mathbf{M}$ look quite different at first glance.
However, it is possible to reconcile the differences by redefining these quantities. In the
Mazur version they are given as exact definitions from the electronic structure. In Dreyer's
version they are to some extent arbitrary quantities, chosen such that the equations become
consistent, and fixed in the end by the closure, so that there is no underlying fixed definition.
We can therefore match the two versions by setting
\begin{align}
  \mathbf{P}^{\text{Dreyer}} &= \mathbf{P}^{\text{Mazur}} - \div\mathbf{Q}^{\text{Mazur}},
  \label{eq:Pmatch}\\
  \mathbf{M}^{\text{Dreyer}} &= \mathbf{M}^{\text{Mazur}}
   - \mathbf{v}\times\big(\mathbf{P}^{\text{Mazur}}-\div\mathbf{Q}^{\text{Mazur}}\big)
   - \Big\langle \textstyle\sum_k\mathbf{v}_k'\times
     \big(\bm{\mu}^{\mathrm{el}}_k-\div\mathbf{Q}_k\big)
     \delta(\Rk_k-\mathbf{R})f\Big\rangle,
  \label{eq:Mmatch}
\end{align}
or equivalently
\begin{equation}
  \Mlor^{\text{Dreyer}}
  = \mathbf{M}^{\text{Mazur}}
  {}- \Big\langle \textstyle\sum_k\mathbf{v}_k'\times
     \big(\bm{\mu}^{\mathrm{el}}_k-\div\mathbf{Q}_k\big)\delta(\Rk_k-\mathbf{R})f\Big\rangle,
  \label{eq:Mlor-match}
\end{equation}
so the Lorentz magnetisation of Dreyer is Mazur's magnetisation up to the correlation of the
particle velocity fluctuations with the dipole and quadrupole moments.
Thus, with a redefinition of polarisation and magnetisation, also these equations can be
brought into a structurally equivalent form. Dreyer then goes on to find constitutive equations
for $\mathbf{P}$ and $\Mlor$ without referring to the electronic definitions, so it makes sense
to check the final constitutive equations for consistency with the electronic expressions in
the end.

The convective quantity $\Mlor$ for magnetization, which Dreyer introduces so that
the magnetisation transforms correctly under Galilean transformations, is, up to the
individual particle-velocity corrections, exactly the form of the magnetisation that appears
naturally in Mazur's formalism. Regarding the
polarisation, it becomes clear that $\mathbf{P}^{\text{Dreyer}}$ differs from the typical
notation in the literature, where $\mathbf{P}$ often denotes the dipole part of the polarisation
as it is the case with $\mathbf{P}^{\text{Mazur}}$; instead it represents the full polarisation
including the quadrupole contribution and possibly higher order as well.

\subsection{Field-energy versus matter-energy representation}

A central source of confusion in the literature is the existence of two equivalent but
superficially different representations of the electromagnetic energy.

In the \emph{field-energy representation} the electromagnetic energy density is the vacuum
expression $u_{\mathrm{em}} = \tfrac12(\varepsilon_0|\mathbf{E}|^2 + \mu_0^{-1}|\mathbf{B}|^2)$,
and the coupling to matter appears entirely in the source terms. In the \emph{matter-energy
representation} part of the electromagnetic energy is absorbed into the matter energy as the
interaction terms $-\Emf\cdot\hat{\mathbf{P}}$ or $\Mlor\cdot\hat{\mathbf{B}}$, and the field energy is
defined through $\bar u_{\mathrm{em}}(\mathbf{D},\mathbf{B})$ with
\begin{equation}
  \frac{\partial\bar u_{\mathrm{em}}}{\partial\mathbf{D}} = \mathbf{E},
  \qquad
  \frac{\partial\bar u_{\mathrm{em}}}{\partial\mathbf{B}} = \mathbf{H}.
  \label{eq:matter-energy-rep}
\end{equation}
Neither representation is more physical than the other. Both are consistent groupings of the
same microscopic energy balance. The total energy of matter plus field is
uniquely defined and conserved, but its splitting between the subsystems is not.
The same freedom is behind the long dispute about the electromagnetic momentum in matter,
which is resolved in the same way \cite{a1:barnett2010}.

We remark that the term $\Mlor\cdot\hat{\mathbf{B}}$ in Dreyer's energy ansatz does not need to be set
externally via an energy functional; it appears intrinsically when the energy equation is
transferred to the relevant electromagnetic quantities $\mathbf{P}$ and $\Mlor$, see above.
In the theory of Dreyer no specific definition of electromagnetic energy is needed.

\subsection{Mazur's polarisation balance versus Dreyer's kinematic identity}

Mazur's \eqref{eq:jP-mazur} and Dreyer's \eqref{eq:kinematic-identity} for the internal electric current densities
are objects of a different kind. Equation \eqref{eq:kinematic-identity} is a \emph{kinematic identity} and introduces polarization and magnetization.
 It is the general
solution of the charge conservation law and contains no material information whatsoever.
$\mathbf{P}$ and $\Mlor$ are potentials for $n_P$ and $\mathbf{J}_P$. Equation
\eqref{eq:jP-mazur}, by contrast, is an \emph{exact microscopic balance}: it says what the bound
current really is in terms of the electronic structure. Consequently, \eqref{eq:jP-mazur}
contains unresolved microscopic fluxes, the terms with $\mathbf{v}_k'$, which
\eqref{eq:kinematic-identity} does not, because they do not have to.

The theories can be reconciled by realizing the different definitions of polarization and
magnetization. As phenomenological theory the equation of Dreyer allows to view the variables
$\mathbf{P}$ and $\Mlor$ much broader than the precise microscopic definition of Mazur.
Indeed, as discussed above the magnetization of Dreyer includes all microscopic processes
which lead to a magnetization of the material including microscopic fluxes.
The two theories are therefore not in conflict, but have different scopes.

\subsection{Velocity-gradient and vorticity terms in the evolution equations}

The evolution \eqref{eq:P-evolution} for $\mathbf{P}$ carries the term
$-\tfrac12(\nabla\mathbf{v}-\nabla\mathbf{v}^{T})\cdot\mathbf{P}$
in the Jaumann time derivative,
i.e.\ a coupling of the polarisation to the velocity vorticity. Nothing of this kind is present
in Mazur. In fact, the definition of the polarization in \eqref{eq:Pdef} can be time-differentiated, which
allows to derive a formal evolution equation of the polarization based on the microscopic picture,
\begin{equation}
  \partial_t\mathbf{P}
  + \div\Big\langle \textstyle\sum_k \dot\Rk_k\,\bm{\mu}^{\mathrm{el}}_k\,\delta(\Rk_k-\mathbf{R})f\Big\rangle
  = \Big\langle \textstyle\sum_k \dot{\bm{\mu}}^{\mathrm{el}}_k\,\delta(\Rk_k-\mathbf{R})f\Big\rangle .
  \label{eq:P-evolution-micro}
\end{equation}
Splitting $\dot\Rk_k=\mathbf{v}+\mathbf{v}_k'$ turns the flux into $\div(\mathbf{v}\otimes\mathbf{P})$
plus a fluctuation term. No contraction of $\nabla\mathbf{v}$ with $\mathbf{P}$ shows up.
This equation is not closed, instead it contains higher order moments that would need closure relations
for the evolution to become solvable. However, no velocity gradients are present in this evolution.
The reason is not a disagreement about physics but a difference in the level of
description.

Mazur works with exact lab-frame ensemble averages. An exact average of an exact microscopic
equation is automatically consistent with any frame change, due to the microscopic dynamics and
there is no need to enforce anything. The velocity-gradient terms are hidden inside the
averages $\avg{\dots\mathbf{v}_k'\dots}$ and are simply not made explicit. A closure, however,
could very well reveal such terms.

Dreyer et al., on the other hand, must \emph{postulate} a constitutive law for a quantity that
they have not defined microscopically. A constitutive law must be frame-indifferent, and the
only frame-indifferent rate of a vector field in a moving continuum is the corotational
(Jaumann) rate, or one of its relatives. Objectivity therefore produces the vorticity term, and
the transport of the bound-charge density produces the $\div\mathbf{v}$ and
$(\mathbf{P}\cdot\nabla)\mathbf{v}$ terms. Hence, the terms appear in Dreyer because a
closure is being made.

\section{Discussion and Outlook}\label{sec:discussion}

\subsection{Summary}

We derived the conservation laws of mass, momentum and energy for a system of charged and
polarisable particles from Mazur's microscopic setup, and we compared them with the bulk
equations of Dreyer, Guhlke and M\"uller. The two sets can be made to coincide. What is needed
is a suitable definition of internal energy, stress and heat flux, together with the
redefinition \eqref{eq:Pmatch} and \eqref{eq:Mmatch} of polarisation and magnetisation.

One difference stays. The field fluctuations contribute
the momentum $\avg{\varepsilon_0\mathbf{e}'\times\mathbf{b}'f}$, which no macroscopic theory can
produce. We have no estimate of its size yet, so we cannot say whether it matters in practice.
The mass corrections, if they are kept, enter at the same place. They are harmless because the
same term also appears in the continuity equation, so the Mazur based system stays consistent
once the total kinetic momentum is defined to include them. All other deviations fit into
a redefinition of stress tensor $\Stot$, heat flux $\mathbf{q}$ and internal energy $u_{\mathrm{int}}$. In the microscopic route these terms
\emph{are} the stress, the heat flux and the internal energy. In the macroscopic route the same
quantities are postulated and then closed.

We emphasise again that the source of the internal energy balance is $\mathbf{J}_e\cdot\Emf$. It pairs the
non-convective current with the electromotive intensity, and only this pairing is invariant. 
Similarly, the electromotive intensity $\Emf$ and the Lorentz magnetisation $\Mlor$ appear by
themselves in the energy balance and in the entropy production. They are not modelling choices.
Going beyond Dreyer et al.\ we showed that the signs of the bound-current ansatz follow from the entropy principle
and are not arbitrary. Ultimately, where the electromagnetic energy sits is a matter of convention. The total energy of
matter plus field is unique and conserved.

Both routes reach the same structure from opposite ends. Mazur starts microscopically and
averages. Dreyer et al.\ start from axioms and close with the entropy principle. The second
route is more general because it also covers mixtures, surfaces and reactions without any
microscopic input. The first route gives the exact expressions. That two such different theories
agree up to higher order terms gives confidence in both.

\subsection{Open issues}

Several questions remain open and are not pursued here.
\begin{itemize}
\item \emph{Moving and deforming media.} The corrected relations \eqref{eq:Dmazur} and
\eqref{eq:Hmazur} contain velocity terms and quadrupole terms whose consequences for the closure
have not been worked out. A complete analysis of all transformation properties is needed
\cite{a1:maugin1988,a1:mueller2022}.
\item \emph{Magnitude of the fluctuation term.} An estimate of
$\avg{\varepsilon_0\mathbf{e}'\times\mathbf{b}'f}$, analogous to the mass-correction estimate of
Sec.~\ref{sec:mass-estimate}, would tell us whether the one remaining difference between the two
theories matters in practice.
\item \emph{Relativistic corrections.} The framework here is non-relativistic. For highly
conducting materials or strong fields, Lorentz-covariant corrections may become important
\cite{a1:jackson1999}.
\item \emph{Surfaces.} The framework of Dreyer et al.\ includes surface balances. Deriving their
microscopic foundation, in the way Mazur does it for the bulk, is an important open problem and
was deliberately left out of the present paper.
\end{itemize}

\section*{Acknowledgements}
Funding by German Research Foundation (DFG) under Research Unit FOR5409: ``Structure-Preserving
Numerical Methods for Bulk- and Interface-Coupling of Heterogeneous Models (SNuBIC)'' (grant \#463312734).
\\[2mm]
Generative AI tools were used to support the preparation of this paper, including for tasks such as phrasing,
consistency checks, and language editing. The authors have critically reviewed all AI-assisted material, they are the
authors of the scientific content, and take full individual responsibility for the results.

\appendix

\section{Microscopic Derivation of the Conservation Laws}\label{app:A}

\subsection{Phase-space setup and the prototype equation}\label{app:A1}

In Mazur's notation we start with \eqref{eq:liouville}, which describes the conservation of
probability in phase space. We now look at $\delta(\Rk_{ki}-\mathbf{R})f$:
\begin{equation}
  \frac{\dd}{\dd t}\big(\delta(\Rk_{ki}-\mathbf{R})f\big)
  = -f(\dot\Rk_{ki}\cdot\nabla)\delta(\Rk_{ki}-\mathbf{R})
  + \delta(\Rk_{ki}-\mathbf{R})\underbrace{\frac{\dd f}{\dd t}}_{=0},
\end{equation}
where we have used $\nabla_{\Rk_{ki}} = -\nabla_{\mathbf{R}}\equiv-\nabla$. Thus
\begin{equation}
  \partial_t\big(\delta(\Rk_{ki}-\mathbf{R})f\big)
  + \div\big(\dot\Rk_{ki}\delta(\Rk_{ki}-\mathbf{R})f\big) = 0 .
  \label{eq:prototype}
\end{equation}
This is the ``prototype'' for our conservation equations. We obtain the explicit conservation
equations of mass, momentum and energy by multiplying this equation with the corresponding
expressions. In the single-particle picture, mass, kinetic momentum and kinetic energy of each
charge carrier are
\begin{equation}
  \rho_{ki} = m_{ki}\delta(\Rk_{ki}-\mathbf{R}),\quad
  \mathbf{p}_{\mathrm{kin},ki} = m_{ki}\dot\Rk_{ki}\delta(\Rk_{ki}-\mathbf{R}),\quad
  e_{\mathrm{kin},ki} = \tfrac12 m_{ki}\dot\Rk_{ki}^2\delta(\Rk_{ki}-\mathbf{R}).
\end{equation}
We therefore multiply \eqref{eq:prototype} by $m_{ki}$, $m_{ki}\dot\Rk_{ki}$ and
$\tfrac12 m_{ki}\dot\Rk_{ki}^2$, sum over all $k$ and $i$, rearrange, and finally take the
ensemble average to switch to the macroscopic picture.

\subsection{Continuity equation}\label{app:A2}

Multiplying \eqref{eq:prototype} by $m_{ki}$ and summing gives
\begin{equation}
  \partial_t\Big(\sum_{k,i}m_{ki}\delta(\Rk_{ki}-\mathbf{R})f\Big)
  + \div\Big(\sum_{k,i}m_{ki}\dot\Rk_{ki}\delta(\Rk_{ki}-\mathbf{R})f\Big) = 0 .
\end{equation}
The ``level of microscopy'' in this equation is that of single charges. In the thermodynamic
framework, atoms and molecules move together as a unit, which is expressed by
$|\rk_{ki}|\ll|\Rk_k|$. We expand in this smallness up to second order:
\begin{align}
  \sum_{k,i}m_{ki}\delta(\Rk_{ki}-\mathbf{R})
  &\approx \sum_{k,i}m_{ki}\Big(1-\rk_{ki}\cdot\nabla
   + \tfrac12\rk_{ki}\rk_{ki}:\nabla\nabla\Big)\delta(\Rk_k-\mathbf{R}) \nonumber\\
  &= \sum_k m_k\delta(\Rk_k-\mathbf{R})
   + \sum_{k,i}\tfrac12 m_{ki}\rk_{ki}\rk_{ki}:\nabla\nabla\,\delta(\Rk_k-\mathbf{R}),
\end{align}
where the first-order term drops out by \eqref{eq:rki}. For the flux term,
\begin{equation}
\begin{aligned}
  \sum_{k,i}m_{ki}\dot\Rk_{ki}\delta(\Rk_{ki}-\mathbf{R})
  &\approx \sum_k m_k\dot\Rk_k\delta(\Rk_k-\mathbf{R})
  \\
  &\quad + \sum_{k,i}\Big(\tfrac12 m_{ki}\dot\Rk_k\rk_{ki}\rk_{ki}:\nabla\nabla
   - m_{ki}\dot\rk_{ki}(\rk_{ki}\!\cdot\!\nabla)\Big)\delta(\Rk_k-\mathbf{R}),
\end{aligned}
\end{equation}
and again all first-order contributions vanish. Taking the ensemble average and introducing the
macroscopic mass density $\rho$ and velocity $\mathbf{v}$ we obtain \eqref{eq:continuity-full}.
Neglecting the second-order
mass corrections, see Sec.~\ref{sec:mass-estimate}, we arrive at the usual continuity equation
$\partial_t\rho + \div(\rho\mathbf{v}) = 0$.

\subsection{Momentum conservation and the Maxwell stress}\label{app:A3}

Multiplying \eqref{eq:prototype} with $m_{ki}\dot\Rk_{ki}$ and summing gives
\begin{equation}
  \partial_t\Big(\sum_{k,i}m_{ki}\dot\Rk_{ki}\delta f\Big)
  + \div\Big(\sum_{k,i}m_{ki}\dot\Rk_{ki}\dot\Rk_{ki}\delta f\Big)
  = \sum_{k,i}\underbrace{m_{ki}\ddot\Rk_{ki}\delta}_{\text{force on }ki}f
  \equiv \sum_{k,i}\mathbf{F}_{ki}f ,
\end{equation}
where $\delta\equiv\delta(\Rk_{ki}-\mathbf{R})$. The expansion of the first term is already
known; for the second one finds
\begin{multline}
  \sum_{k,i}m_{ki}\dot\Rk_{ki}\dot\Rk_{ki}\delta(\Rk_{ki}-\mathbf{R})
  \approx \sum_k m_k\dot\Rk_k\dot\Rk_k\delta(\Rk_k-\mathbf{R}) \\
  + \sum_{k,i}m_{ki}\Big(\dot\rk_{ki}\dot\rk_{ki}
  + \tfrac12\dot\Rk_k\dot\Rk_k\rk_{ki}\rk_{ki}:\nabla\nabla
  - (\dot\Rk_k\dot\rk_{ki}+\dot\rk_{ki}\dot\Rk_k)(\rk_{ki}\cdot\nabla)\Big)
  \delta(\Rk_k-\mathbf{R}),
\end{multline}
where again all first-order contributions vanish. The right-hand side is the sum of the forces
on each single charge carrier, i.e.\ the Lorentz force. Inserting the microscopic Maxwell
equations \eqref{eq:micro-maxwell} gives the familiar rearrangement
\begin{equation}
  \sum_{k,i}e_{ki}(\mathbf{e}+\dot\Rk_{ki}\times\mathbf{b})\delta(\Rk_{ki}-\mathbf{R})
  = \div\Big[\varepsilon_0\mathbf{e}\mathbf{e} + \frac{1}{\mu_0}\mathbf{b}\mathbf{b}
   - \frac12\Big(\varepsilon_0\mathbf{e}^2+\frac{1}{\mu_0}\mathbf{b}^2\Big)\Iden\Big]
   - \varepsilon_0\partial_t(\mathbf{e}\times\mathbf{b}) .
\end{equation}
Taking the ensemble average, splitting the velocities according to \eqref{eq:vkprime} so that
$\avg{\sum_k m_k\mathbf{v}_k'\delta f}=0$, and splitting the fields into an averaged and a
fluctuating part,
\begin{equation}
  \mathbf{e}' \equiv \mathbf{e}-\avg{\mathbf{e}} = \mathbf{e}-\mathbf{E},
  \qquad
  \mathbf{b}' \equiv \mathbf{b}-\avg{\mathbf{b}} = \mathbf{b}-\mathbf{B},
  \label{eq:field-fluctuations}
\end{equation}
the momentum conservation equation becomes \eqref{eq:momentum-full},
where $\Stot$ contains the contributions both from the velocities inside the macroscopic
particles, related to $\rk_{ki}$, and from the deviations of the peculiar velocities from the
mean velocity, as well as the deviations of the EM fields from their averages:
\begin{multline}
  \Stot = -\Big\langle\sum_k m_k\mathbf{v}_k'\mathbf{v}_k'\delta(\Rk_k-\mathbf{R})f
  + \Big[\frac12\Big(\varepsilon_0\mathbf{e}'^2+\frac{1}{\mu_0}\mathbf{b}'^2\Big)\Iden
  - \Big(\varepsilon_0\mathbf{e}'\mathbf{e}'
  + \frac{1}{\mu_0}\mathbf{b}'\mathbf{b}'\Big)\Big]f \\
  + \sum_{k,i}m_{ki}\Big(\dot\rk_{ki}\dot\rk_{ki}
  + \tfrac12\dot\Rk_k\dot\Rk_k\rk_{ki}\rk_{ki}:\nabla\nabla
  - (\dot\Rk_k\dot\rk_{ki}+\dot\rk_{ki}\dot\Rk_k)(\rk_{ki}\cdot\nabla)\Big)
  \delta(\Rk_k-\mathbf{R})f\Big\rangle .
  \label{eq:sigma-micro}
\end{multline}
The Maxwell stress tensor thus emerges by itself; it is not postulated.

\subsection{Energy conservation and the Poynting vector}\label{app:A4}

Multiplying \eqref{eq:prototype} with $\tfrac12 m_{ki}\dot\Rk_{ki}^2$ and summing gives
\begin{equation}
  \partial_t\Big(\textstyle\sum_{k,i}\tfrac12 m_{ki}\dot\Rk_{ki}^2\delta f\Big)
  + \div\Big(\textstyle\sum_{k,i}\tfrac12 m_{ki}\dot\Rk_{ki}^2\dot\Rk_{ki}\delta f\Big)
  = \textstyle\sum_{k,i}W_{ki}f,
\end{equation}

with $W_{ki}= e_{ki}\big(\mathbf{e}+\dot\Rk_{ki}\times\mathbf{b}\big)\cdot\dot\Rk_{ki}\,\delta$ the work done on the
carrier $ki$. For the right-hand side one finds, again with the Maxwell equations,
\begin{equation}
  \textstyle\sum_{k,i}e_{ki}\dot\Rk_{ki}\delta(\Rk_{ki}-\mathbf{R})\cdot\mathbf{e}
  = -\partial_t\Big(\frac{\varepsilon_0}{2}\mathbf{e}^2
   + \frac{1}{2\mu_0}\mathbf{b}^2\Big)
   - \frac{1}{\mu_0}\div(\mathbf{e}\times\mathbf{b}) .
\end{equation}
For the energy equation we can exploit a simple trick: since for the mass terms we only want to
consider the motion of whole particles, we put all smaller contributions into the internal
energy $u_{\mathrm{int},m}$, and the part of the flux which contains inner-particle motion into
the heat flux $q_m$. Explicitly,
\begin{align}
  u_{\mathrm{int},m} &= \sum_{k,i}\tfrac12 m_{ki}\Big(\dot\rk_{ki}^2
   - 2(\dot\rk_{ki}\cdot\dot\Rk_k)(\rk_{ki}\cdot\nabla)
   + \tfrac12\dot\Rk_k^2\,\rk_{ki}\rk_{ki}:\nabla\nabla\Big)\delta(\Rk_k-\mathbf{R}), \\
  q_m &= \sum_{k,i}\tfrac12 m_{ki}\Big(\dot\rk_{ki}^2\dot\Rk_k
   + 2(\dot\rk_{ki}\cdot\dot\Rk_k)\dot\rk_{ki}
   + \tfrac12\dot\Rk_k^2\dot\Rk_k(\rk_{ki}\rk_{ki}:\nabla\nabla) \nonumber\\
  &\hspace{2.2cm} - \big(\dot\Rk_k^2\dot\rk_{ki}
   + 2(\dot\rk_{ki}\cdot\dot\Rk_k)\dot\Rk_k\big)(\rk_{ki}\cdot\nabla)\Big)
   \delta(\Rk_k-\mathbf{R}),
\end{align}
and with this notation the energy conservation equation reads
\begin{multline}
  \partial_t\Big[\tfrac12\sum_k m_k\dot\Rk_k^2\delta(\Rk_k-\mathbf{R}) + u_{\mathrm{int},m}
  + \frac{\varepsilon_0}{2}\mathbf{e}^2 + \frac{1}{2\mu_0}\mathbf{b}^2\Big]
  \\[2pt]
  + \div\Big[\tfrac12\sum_k m_k\dot\Rk_k^2\dot\Rk_k\delta(\Rk_k-\mathbf{R}) + q_m
  + \frac{1}{\mu_0}\mathbf{e}\times\mathbf{b}\Big] = 0 .
\end{multline}
The Poynting vector representing the flow of EM energy is thus
\begin{equation}
  \mathbf{S} = \frac{1}{\mu_0}\mathbf{e}\times\mathbf{b} .
  \label{eq:poynting}
\end{equation}
Note that the EM momentum density is equal to $c^{-2}\mathbf{S}$, so that the law stated by
Feynman \cite{a1:feynman} is obeyed: any flow of energy through a unit area per unit time,
multiplied by $c^{-2}$, must equal the momentum per unit volume in this space. This is the
requirement which several macroscopic couplings in the literature violate.

Taking the ensemble average, splitting off the deviations from the mean velocity and the field
fluctuations \eqref{eq:field-fluctuations}, and putting them into the internal energy and the
heat flux, we obtain \eqref{eq:energy-full} with
\begin{align}
  u_{\mathrm{int}} &= \Big\langle \sum_{k,i}\tfrac12 m_{ki}\Big(\dot\rk_{ki}^2
   - 2(\dot\rk_{ki}\cdot\dot\Rk_k)(\rk_{ki}\cdot\nabla)
   + \tfrac12\dot\Rk_k^2\rk_{ki}\rk_{ki}:\nabla\nabla\Big)\delta f \nonumber\\
  &\hspace{1.2cm} + \sum_k\tfrac12 m_k(\mathbf{v}_k')^2\delta f
   + \frac12\Big(\varepsilon_0\mathbf{e}'^2+\frac{1}{\mu_0}\mathbf{b}'^2\Big)f\Big\rangle,
  \label{eq:uint-micro}\\
  \mathbf{q} &= -\big(\tfrac12\rho v^2 + u_{\mathrm{int}}\big)\mathbf{v}
   + \Big\langle\sum_k\tfrac12 m_k\dot\Rk_k^2\dot\Rk_k\delta f\Big\rangle
   + \avg{q_m f}
   + \frac{1}{\mu_0}\avg{\mathbf{e}'\times\mathbf{b}'f} .
  \label{eq:q-micro}
\end{align}
It is possible to simplify $\mathbf{q}$ further by inserting $u_{\mathrm{int}}$ and the
definition of $\mathbf{v}_k'$.

\bibliographystyle{plainnat}
\bibliography{TD_EM_coupling}

\end{document}